\documentclass[letterpaper,12pt]{JHEP3}
\usepackage{amssymb}
\usepackage{mathrsfs}
\usepackage{cite}
\usepackage{texdraw}
\usepackage[english]{babel}
\usepackage{epsfig}

\newcommand{\be}{\begin{equation}}
\newcommand{\ee}{\end{equation}}
\newcommand{\bea}{\begin{eqnarray}}
\newcommand{\eea}{\end{eqnarray}}
\newcommand{\del}{\partial}
\newcommand{\f}{\frac}
\newcommand{\e}{\epsilon}

\newcommand{\h}{\hat}
\newcommand{\g}{\gamma}

\newcommand{\hs}[1]{\hspace{#1 mm}}

\let\bm=\bibitem
\def\fs{|\phi|^2}
\def\del{\partial}
\let\la=\label
\def\nn{\nonumber}

\newcommand{\w}[1]{\\[0.#1cm]}
 \def\det{{\rm det\,}}

\def\a{\alpha}

\def\c{\gamma}

\def\d{\delta}

\def\e{\epsilon}
\def\ve{\varepsilon}
\def\f{\phi}
\def\F{\Phi}

\def\k{\kappa}
\def\l{\lambda}

\def\m{\mu}
\def\n{\nu}
\def\r{\rho}

\def\th{\theta}

\def\O{\Omega}
\def\o{\omega}

\def\lra{\leftrightarrow}

\def\g{1+\e |\phi|^2}

\def\fs{|\phi|^2}
\def\ra{\rightarrow}
\def\lra{\leftrightarrow}
\def\qq{\quad\quad}

\title{Supersymmetric Strings and Waves in $D=3, N=2$ Matter Coupled 
Gauged Supergravities}
\author{Nihat Sadik Deger \\ Feza G\"{u}rsey 
Institute\\ 
Emek Mah. No:68, Cengelkoy \\
81220, Istanbul, Turkey \\
E-mail: \email{deger@gursey.gov.tr}}

\author{\"{O}zg\"{u}r Sar{\i}o\u{g}lu \\ 
Department of Physics \\
Middle East Technical University \\ 
06531, Ankara, Turkey \\
E-mail: \email{sarioglu@metu.edu.tr}}

\abstract
{We construct new 1/2 supersymmetric solutions 
in $D=3, N=2$, matter  
coupled, U(1) gauged supergravities and study some of their properties. 
In the most general case they represent a 
string superposed with gravitational and 
Chern-Simons electromagnetic waves. The waves are attached to  
the string and 
the solution satisfies an electromagnetic self-duality relation. When the 
sigma model is non-compact it interpolates between an 
asymptotically Kaigorodov 
space and a naked singularity. For the compact sigma model  
there is a regular horizon with 
the Kaigorodov geometry and  
asymptotically it is either Minkowskian or a 
pp-wave. When the sigma manifold is flat our solutions describe
either $AdS_3$ or Kaigorodov space or a pp-wave in $AdS_3$.}

\keywords{Supergravity models, $AdS/CFT$ correspondence, pp-waves}

\preprint{}

\begin{document}

\section{Introduction}

Supersymmetric solutions of supergravity theories have played a major role 
in 
many of the recent advances in string/M theory. Among supergravity 
theories interest in the gauged ones increased dramatically after the 
advent of the $AdS/CFT$ duality \cite{mal1,mal2,mal3} since 
generically they possess scalar fields with potentials which have
some $AdS$ extrema. To test this conjecture, gravity theories 
in three dimensional anti-de Sitter space (for a review see 
\cite{nicolai}) is especially appropriate 
because they are claimed to be related to two dimensional $CFT$'s and such 
conformal field theories are the best understood ones.

It is clearly desirable to go beyond the supergravity approximation in 
$AdS/CFT$ correspondence and recently it is understood that 
pp-wave backgrounds provide such an opportunity \cite{bmn}. Plane waves 
(which are a subset of pp-waves) 
can arise by taking the Penrose-G\"{u}ven 
limit \cite{penrose, guven} of $AdS_p \times S^q$ backgrounds 
\cite{papa1} and string theory in many of them turns out to be 
exactly solvable \cite{met1} (for an up-to-date review and more references 
see \cite{sadri}). A way to improve our understanding of such spacetimes 
is to construct asymptotically pp-wave solutions. Such black strings  
were recently studied in \cite{hub1, hub2, has1, has2}.

Motivated by these, in this paper we find new superysmmetric
solutions in the matter coupled $D=3, N=2$, U(1) gauged supergravities and
study some of their properties. This model was constructed 
in \cite{ads2}
and admits both compact and non-compact sigma model manifolds.  
There is also a well-defined flat sigma model limit. The theory contains 
only a Chern-Simons gauge field and no Maxwell term. 
The only known supersymmetric solutions of this model are static,
uncharged strings \cite{ads2} and vortices \cite{sam1}. Among our 
solutions we have stationary and charged generalizations of the strings
given in \cite{ads2}. Moreover, by carefully analyzing Killing spinor and
field equations we show that it is possible to add two types of waves to
these strings. Some time ago Garfinkle and Vachaspati developed a 
technique
\cite{vac, gar} which allows one to introduce waves to an
already constructed solution with a null Killing vector. Our first type of
wave turns out to be exactly the one obtained by this method starting from
the string solution. The second type exists only when there is a
non-trivial radial dependence in the Chern-Simons vector field and
therefore can be regarded as an electromagnetic wave. It is not possible
to separate the waves from the string and the solution satisfies 
an electromagnetic self-duality relation. The charge of the 
solution can 
be set to zero by choosing the scalar field and the Killing spinor real.

The scalar fields are not affected with these
additional waves but the global properties of the metric are quite 
sensitive to them.
For example when the sigma model manifold is non-compact, asymptotic
geometry for the solution without waves approaches to $AdS_3$ whereas with
waves it becomes the Kaigorodov space \cite{kaig}. This is  
a homogeneous Einstein space which describes a pp-wave in $AdS$ and in 
three dimensions it is equivalent to the 
extremal BTZ black hole \cite{btz}.
This space has
already appeared in the $AdS/CFT$ context \cite{pope,real} where a duality
between a supergravity theory in the Kaigorodov space and a CFT living on
its boundary was proposed (see also \cite{podol, kaig2, kaig3, lu}). This 
boundary is 
related to the usual $AdS$
boundary by an infinite Lorentz boost \cite{pope}. 
For the compact sigma 
model our
solutions exhibit an event horizon 
whose geometry is deformed from $AdS$ to Kaigorodov space by the presence 
of the waves as was observed for $M2, M5$ and $D3$ 
branes in \cite{pope}. All curvature invariants are finite at 
the horizon and we show that geodesics never cross it.
When the electromagnetic wave is
absent the asymptotic geometry is Minkowskian but it is a pp-wave
otherwise. For the flat sigma model our solutions describe 
either $AdS_3$ or Kaigorodov space or a pp-wave in $AdS_3$.

The plan of this paper is as follows. In section 2 we begin with a review 
of the $N=2$ gauged supergravity with matter. In section 3 we derive
our supersymmetric string and wave solutions. We conclude in section 4 
with some comments and future directions. A brief 
introduction to the Garfinkle-Vachaspati solution generating method 
together with its application to our string solution is 
given in the appendix.

\section{The Model}

In this paper we consider $N=2$, U(1) gauged supergravity in $D=3$ 
interacting 
with an arbitrary number of matter multiplets which was constructed in 
\cite{ads2} 
using Noether's procedure. Its higher dimensional origin is yet to be 
discovered. The
boundary symmetries of this theory were studied in \cite{mc2} and its
extension by including a Fayet-Iliopoulos term was given in \cite{sam1}.
Holographic RG flows in this model were analyzed in \cite{rg}. There is
another $N=2$ theory constructed in \cite{it} where scalars are not
charged with respect to the U(1) $R$-symmetry group. Therefore in 
\cite{it}
there is only a cosmological constant and no scalar potential. The
connection between these two models for the flat sigma model was described
in \cite{mc2}.  Let us also mention that the model we consider in this
paper \cite{ads2} is a member of a class of theories called abelian
Chern-Simons Higgs models coupled to gravity (see \cite{sam1, verbin} and
references therein). The field content of the theory is:
\

\, $\bullet$ The supergravity multiplet: \{$e_\m{}^a$, $ \psi_\m$, 
$A_\m$\}

\, $\bullet$ The scalar multiplet ($K$ copies): \{$\phi^\a$, $\l^r$\}
\

\noindent
All fields except the graviton $e_\m{}^a$ and the gauge field $A_\m$ are 
complex. In \cite{ads2}, the following sigma model manifolds $M$ were 
considered:
\be
M_+= CP^K={SU(K+1)\over SU(K)\times U(1)}\ , \quad\quad M_-= CH^K=
{SU(K,1)\over SU(K)\times U(1)}\ .
\la{pm2}
\ee
Note that $U(1)$ is the $R$-symmetry group.
We define the parameter $\e=\pm$1 to indicate the manifolds
$M_\pm$.
In this paper we choose $K=1$  and consider the following cases, 
$S^2=SU(2)/U(1)$ and
$H^2=SU(1,1)/U(1)$.
The bosonic part of the Lagrangian is
\footnote{Our conventions are as 
follows: We take $\eta_{ab} = (-,+,+)$ and 
$\e^{\m\n\r}=\sqrt{-g}\c^{\m\n\r}$. 
In coordinate basis a convenient representation for 
$\gamma^{a}$ 
matrices is
$\gamma_0=i\sigma^3, \gamma_1=\sigma^1, \gamma_2=\sigma^2$ 
with 
$\epsilon^{012}=1$. Here 0,1,2 refer to the tangent time, radial and 
theta directions, respectively, and $\c^2$ is the charge conjugation 
matrix.} 
\be
{\cal L} = \sqrt{-g}\left(
{1\over 4} R
-{1\over 16ma^4}\, {\e^{\m\n\r} \over \sqrt{-g}} A_\m \del_\n A_\r 
- { |D_\m\f|^2 \over a^2(1+\e |\f|^2)^2} 
-V(\f) 
\right)\, ,
\la{ba1}
\ee
where $D_\m\f=(\del_\m -i\e A_\m)\f$ \, and
the potential is given by
\be
V(\f)= 4m^2a^2C^2\left( |S|^2-{1\over 2a^2}C^2\right)\, .
\la{a2}
\ee
Functions $C$ and $S$ are defined as 
\be
C= {1- \e\fs \over \g}\ , \hs{5} S = {2\f \over\g}\ .
\la{cs2}
\ee
\FIGURE{
\centerline{\epsfxsize=3.3truein
\epsffile{fig1.eps}
\hspace{0.4in}
\epsfxsize=3.3truein
\epsffile{fig2.eps}
}
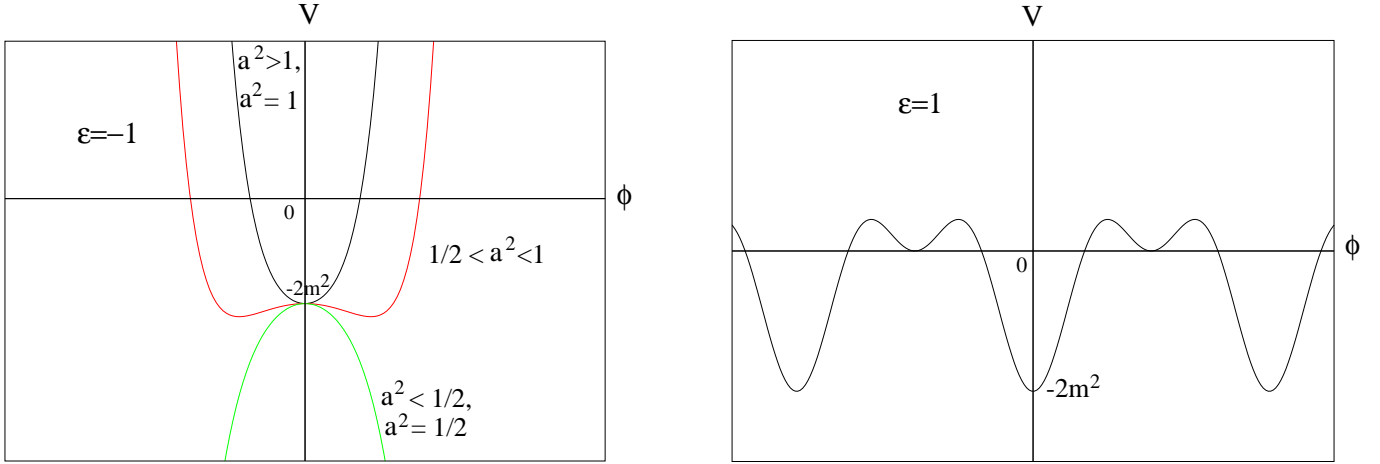
\caption{The scalar potential $V$ plotted with respect to $\phi$.}
}

Note that the following algebraic relations hold:
\be
|\f|^2={\e (1-C) \over (1+C)} \, , \hs{5} \e |S|^2=1-C^2 \, .
\la{algrel}
\ee
The constant $``a"$ is the characteristic
curvature of $M_\pm$ (e.g. $2a$ is the inverse radius in the case
of $M_+=S^2$).
The gravitational coupling constant $\k$ has been
set equal to one and $-2m^2$ is the $AdS_3$
cosmological
constant.
Unlike in a typical $AdS$ supergravity coupled to matter,
the constants $\k,a,m$ are not related to each other for
non-compact scalar manifolds, while $a^2$ is quantized in terms of
$\k$ in the compact case so that ${\k ^2 \over a^2}$ is an
integer \cite{ads2}. When $\e=-1$ for all $a^2$ there is a supersymmetric 
$AdS$ vacuum at $\f=0$ and a non-supersymmetric but stable (it satisfies 
Breitenlohner-Freedman bound \cite{breit}) $AdS$ vacuum for $1/2<a^2<1$. 
When $\e=1$ there are supersymmetric $AdS$, Minkowski and  
non-supersymmetric de Sitter vacua (see figure 1).

The nonlinear scalar covariant derivative $P_\m$ and the $U(1)$ connection
$Q_\m$ are defined as
\bea
P_\m &=& {2 \del_\m\f\over \g}  -i\e  A_\m S\ ,
\nn\w2
Q_\m &=& {i \f \stackrel{\lra}{\del_\m}\f^*  \over \g}\,+A_\m C\, .
\eea

The bosonic field equations that follow from the Lagrangian (\ref{ba1}) are
\bea
R_{\m\n} &=& {1\over a^2} \,P_{(\m} P_{\n)}^*
+ 4 V g_{\m\n} \ ,
\label{e1}
\w2
\e^{\m\n\r}  F_{\n\r} &=&- 4\epsilon im a^2 \sqrt{-g}\,
[P^\m S^* - (P^{\m})^*S]\ ,
\label{e2}
\w2
{1\over \sqrt{-g}} \del_\m \left(\sqrt{-g} g^{\m\n} P_\n\right)
&=& i\e Q_\m P^\m+ 2a^2 \left(1+\e |\f|^2\right)  {\del V\over 
\del\phi^*}\, .
\label{e3}
\eea

The supersymmetric version of the Lagrangian (\ref{ba1}) is invariant 
under the following fermionic supersymmetry transformations 
\bea
\la{susy1}
\d \psi_\m &=& \left(\del_\m + {1\over 4} \o_\m{}^{ab} \c_{ab}
-{i\over 2 a^2}\,Q_\m \right) \ve +m \c_\m C^2 
\ve\ , \w2
\d\l &=&\left(- {1\over 2a} \c^\m P_\m - 2\e m a \, CS \right) \ve\, .
\la{susy2}
\eea
With these preliminaries, we now search for supersymmetric solutions of 
this model in the next section.

\section{Supersymmetric String and Wave Solutions}
Our metric ansatz is
\be
ds^2= -F^2dt^2 + H^2(Gdt+d\theta)^2+ dr^2 ,
\la{metric}
\ee
where $F,G$ and $H$ are functions of $r$ only.
We choose the dreibeins of this metric as
\bea
e_{r1}=1 \,\, , \,\, e_{t0}=F \,\, , \,\, e_{t2}=GH \,\, , \,\, e_{\th 
2}=H \,\, , 
\nn \\ 
e^r_{\,\, 1}=1 \,\, , \,\, e^t_{\,\, 0}= -\frac{1}{F} \,\, , \,\, e^\th 
_{\,\, 
0}=\frac{G}{F} \,\, , \,\, e^\th _{\,\, 2}=\frac{1}{H} \,\, ,
\eea
and the connection 1-forms turn out to be
\bea
\o_t^{\,\, 01}=-F' + \frac{H^2}{2F}GG' \,\, , \,\, \o_t^{\,\, 12}=-GH' 
-\frac{G'H}{2} \,\, , \nn \\
\o_r^{\,\, 02}=\frac{H}{2F}G' \,\, , \,\, \o_\th ^{\,\, 01} = 
\frac{H^2}{2F}G'
\,\, , \,\, \o_\th ^{\,\, 12}=-H' \,\, ,
\eea
where prime indicates derivative with respect to $r$.
The determinant of the metric is $\sqrt{-g}=FH$ and the nontrivial
components of its inverse are given as 
\be
g^{tt}= -\frac{1}{F^2} \,\, , \,\, g^{\th t}= \frac{G}{F^2} \,\, , \,\,
g^{rr}=1 \,\, , \,\, g^{\th \th}= \frac{1}{H^2} - \frac{G^2}{F^2} \,\, .
\ee

We choose the scalar field to be of the form
\be
\f=R(r)e^{in\theta}e^{ik t}\, , 
\la{ans1}
\ee
where $n$ and $k$ are real constants. For the vector field we pick the 
following gauge 
\be
A_\m=(A_t, A_r, A_\theta)=(\psi(r),0, \chi(r))\, .
\la{ans2}
\ee

All fermions are set to zero. However in order to obtain a supersymmetric 
solution we still need to solve (\ref{susy1}) and (\ref{susy2}). 
Solutions can be divided into two classes: $a^2\neq 0$ and $a^2=0$.

\subsection{Non-linear Sigma Models ($a^2 \neq 0$)}

From 
$\d \l=0$, we find

\be
\left[{i(k-\e \psi) S \over F}\c_0 - \left({G \over F} \c_0  - {1\over 
H}\c_2 
\right)i(n-\e\chi) S + {R'\over R} S\c_1 + 4\e ma^2 CS \right]\ve=0 \, ,
\la{lambda}
\ee
and from $\d\psi_{\m}=0$, we get
\bea
\del_t \ve &=&
\left[{iQ_t \over 2a^2} - \left({F' \over 2} - {H^2 GG' \over 
4F}+mC^2GH\right)\c_2 + \left({G'H \over 4} + {GH' \over 
2} +mC^2F\right)\c_0  
\right]\ve \,\,\,\,\,
\la{psi1}
\w2
\del_{\th} \ve&=& \left[ {iQ_{\th} \over 2a^2} + \left({H^2G' \over 
4F} -mC^2H\right)\c_2 + 
{H' \over 2} \c_0 \right]\ve \, , 
\la{psi2} \w2
\del_r \ve &=& -\left[{HG' \over 4F} + mC^2\right]\c_1\ve \, .
\la{psi3}
\eea

For a 1/2 supersymmetric solution we assume a projection of the form
\be
\c_1\ve= p \, \ve \, , \hs{10} (p^2=1) \, . 
\la{proj}
\ee

The above projection applied to (\ref{lambda}) 
gives (note that $\c_0=-\c_2 
\c_1$)
\bea
\la{scalar}
{R'\over R} &=& - 4p\e ma^2 C \, , \w2
{\e \over F}(\psi -G\chi) &=& -{1\over F}(Gn-k) - {p \over H} (n-\e 
\chi)\, .
\label{vec1}
\eea

Equation (\ref{scalar}) is integrable and the result is
\be
C^2={1 \over 1+4\e R_0^2e^{-8\e pma^2r}} \, \, , 
\ee
where $2R_0$ is an integration constant and can be set to 1 by a shift
in $r$. The $r$ dependence of the 
functions $R$ and $S$ can be obtained using (\ref{algrel}). When $\e=1$ 
the range of the coordinate $r$ is $(-\infty,\infty)$, and when  
$\e=-1$ it is $[0, -p\infty)$. The scalar field is smooth when the 
$r$-coordinate is in these intervals. 
Now using the projection (\ref{proj}) in (\ref{psi1}) and 
(\ref{psi2}), we 
find that the metric functions should obey
\bea
\la{f1}
{H'\over H} &=& {pG'H \over 2F} - 2 p m C^2 \, , \w2
{F'\over F} &=& -{pG'H \over 2F} - 2 p m C^2 \, . 
\la{f2}
\eea

In order to determine the Killing spinor, we begin from (\ref{psi3}) which
fixes its form as
\be
\ve=\sqrt{F} e^{\sigma(t,\theta)} \,,
\ee
where we used (\ref{f2}) too. Inserting this result in (\ref{psi2}) we 
get
\be
{\del \sigma \over {\del \theta}}= {i \over {2a^2}} \left( {2nR^2 \over
{1+\e R^2}} + \chi C \right) \, .
\la{spinor}
\ee
Since the right-hand side of this equation is a function of $r$ only we 
conclude that 
\be
{\del \sigma \over {\del \theta}} = ic_1,
\ee
where $c_1$ is a real constant. A similar analysis of 
(\ref{psi1}) gives
\be
{\del \sigma \over {\del t}} = {i \over {2a^2}} \left( {2kR^2 \over
{1+\e R^2}} + \psi C \right) \, ,
\ee
which implies
\be
{\del \sigma \over {\del t}} = ic_2,
\ee
where $c_2$ is another constant. 
The final expressions for the Killing spinor and vector field components 
are
\bea
\la{vec2}
\ve &=&\sqrt{F} e^{ic_1\th}e^{ic_2t} (p+\c_1)\ve_0 \, , \w2
\la{vec3}
\chi&=& \e n - {\e (n-2\e c_1a^2) \over C} \, , \w2
\psi&=& \e k - {\e (k-2\e c_2a^2) \over C}\, ,
\la{vec4}
\eea
where $\ve_0$ is a constant spinor. Notice that when the spinor and 
the scalar 
field (\ref{ans1}) are real , the vector field vanishes. This is the 
chargeless limit 
of our 
solutions.
Equations (\ref{vec3}) and (\ref{vec4}) together with (\ref{vec1}) 
put a strong restriction on the metric functions
\be
k - 2c_2a^2\e=(n-2c_1a^2\e)\left(G+{pF \over H}\right) \, .
\la{rest}
\ee

This completes our investigation of the
supersymmetry variations. Now we have to check the field equations. The
scalar field equation (\ref{e3}) is identically satisfied. The vector
field equation (\ref{e2}) is trivially satisfied when the free index
$\m=r$ due to the fact that $\mathrm{Im}\{ P^r S^* \}=0$. However when 
$\m=t$ and
$\m=\th$, we get
\bea
\chi'&=& 4\e pma^2 |S|^2 (n-\e\chi)\, , \w2
\psi'&=& 4\e pma^2 |S|^2 (k-\e\psi)\, ,
\eea
where we also used (\ref{vec1}). These two equations have to be consistent
with the supersymmetric forms of $\chi$ and $\psi$ given in 
(\ref{vec3}) and (\ref{vec4}) which is indeed the case.
Finally, after a lengthy calculation Einstein's equations (\ref{e1}) 
can be shown to be satisfied provided that 
the following is true:
\be
p\left({HG'\over F}\right)' + \left({HG'\over F}\right)^2 = 
4mC^2 {HG'\over F} - {2 |S|^2 \over a^2} \left({n-\e \chi 
\over H}\right)^2 \, .
\la{f3}
\ee

To summarize, 1/2 supersymmetry breaking projection 
(\ref{proj}) completely determines the scalar and vector fields. It only 
remains to solve equations 
(\ref{f1}), (\ref{f2}) and (\ref{f3}) with the condition (\ref{rest})
which we do next. Equations (\ref{f1}) and (\ref{f2}) can be used 
to determine the metric 
functions $G$ and $F$ in terms of $H$ as:
\be
F= {f_0|S|^{\e/a^2} \over H}\, , \hs{5} G=-{pF \over H} + g_0\, ,
\la{fh}
\ee
where $f_0$ and $g_0$ are real integration constants. Note that 
$f_0$ can never be zero. In obtaining this, we
used definitions of $C$ and $S$ (\ref{cs2}) together with 
(\ref{scalar})
to write
\be
C'=4pma^2C|S|^2 \, , \hs{5} |S|'=-4\e pma^2C^2|S| \, .
\la{derivative}
\ee

The restriction (\ref{rest}) fixes the constant $g_0$ in (\ref{fh}) as
\be
k - 2c_2a^2\e=(n-2c_1a^2\e)g_0\, \label{ozg1}.
\ee
When $(g_0 \neq 0)$ this relation induces an electromagnetic 
self-duality condition 
\be
\psi - \epsilon k = g_0 (\chi - \epsilon n) \;\;\;\; \mbox{or} \;\;\;\;
E = - g_0 B \, , 
\ee
where we denoted the components of the 
electromagnetic field tensor in the orthonormal basis as $F_{01}=E$ and 
$F_{12}=B$. Now from (\ref{f1}) and (\ref{f2}), we have
\be
{HG'\over F}= p\left( {2H' \over H} + 4pmC^2 \right)\, .
\ee
Inserting this and (\ref{derivative}) in (\ref{f3}), we finally obtain
\be
(H^2)'+4pmC^2H^2 = {(n-2\e c_1a^2)^2 \over 4pma^4 C^2} 
+ c_0 \, ,
\la{final}
\ee
where $c_0$ is a real constant. Now our problem is reduced to a single 
linear, ordinary, first order differential equation for $H^2$. 
The most general solution of (\ref{final}) is
\be
H^2= h_0 |S|^{\e/a^2} + 
h_1 F(-{\e \over 2a^2}, 1; 1- {\e \over 2a^2}; \e |S|^2) + 
{h_2 \over C^2} \, ,
\label{most}
\ee
where $h_0$ and $h_1=[c_0/4pm - h_2(1+2\e a^2)]$ are arbitrary 
real constants and  
\be
h_2 = -{\e (n-2\e c_1a^2)^2 \over 32 m^2 a^6} \, . 
\la{h0}
\ee

Here $F(a,b;c;z)$ is a hypergeometric function\footnote{The 
function $F(a,b;c;z)$ 
satisfies $z(1-z)F''+ [c-(a+b+1)z]F'-abF=0$. It has the property 
$F(1,1;2;z)=-z^{-1}\ln(1-z)$ and $F(\frac{1}{2},1;\frac{3}{2};-z^2)
=z^{-1} $arc$\tan z$.}. Using (\ref{fh}) we can write our 
metric (\ref{metric}) as
\be
ds^2= -2pf_0|S|^{\e /a^2}dvdt + H^2dv^2 + dr^2 \, ,
\hs{5} v \equiv \th + g_0t \, .
\la{metricfinal}
\ee
It is clear that constants $p$ and 
$f_0$ can be discarded by redefining the $t$ coordinate in 
(\ref{metricfinal}). Furthermore,
the magnitude of one of the integration constants in $H^2$ can be set to 
1 by a rescaling of the coordinates $v$ and $t$. 

Let us now comment on each 
term in $H^2$. When $h_1=h_2=0$  using 
(\ref{most}) in (\ref{fh}) 
we see that $G$ is constant.
[Note that $h_2=0$ implies that the vector field is constant too, i.e. 
$\chi=\e n = 2c_1a^2$ and $\psi=\e k = 2c_2 a^2$.]
In this case the metric (\ref{metricfinal}) becomes
\be
ds^2=h_0 |S|^{\e/a^2}\left[-{f_0^2 \over h_0^2} dt^2 + 
\left(d\th +\left[g_0 -{pf_0 \over 
h_0}\right]dt\right)^2 \right] +dr^2\, .
\la{stringmetric}
\ee

When $G=g_0-pf_0/h_0=0$ this solution represents a static 
string. The uncharged version of this solution, i.e. 
$\chi=\psi=0$ or $n=k=c_1=c_2=0$, was studied in \cite{ads2}. 
When $G$ is equal to a non-zero constant this solution can be interpreted as a 
rotating (stationary) string provided that the coordinate $\th$ is 
periodic with
$0\leq \th < 2\pi$. 
Actually this metric can be obtained from the 
static one by a simple boost in the $\th-t$ plane. 
However, this transformation is well-defined only locally, and therefore 
one ends up with a globally stationary solution \cite{stachel}. Despite 
the fact that the vector field is constant for this string solution, 
there is still a non-zero charge since $A_\m$ is a Chern-Simons gauge 
field. Unlike the Maxwell theory,
the charge $Q$ associated with it is obtained by 
integrating $A_\m$ and not its derivative (see e.g. \cite{it} for a 
nice discussion of this). Making a coordinate redefinition
$(\th \ra \th-tk/n)$, the time dependence of the scalar field (\ref{ans1}) 
and the Killing spinor (\ref{vec2}) can be turned off \cite{it}. This also 
sets $A_t=\psi=0$. Then the charge associated with $A_\th=\chi$ is 
\be
Q=2\e n =4c_1a^2 \, .
\ee

Let us emphasize that even when one starts with 
$h_0=0$ in (\ref{most}), by making a coordinate redefinition 
($\tau=t + h_0 v/2pf_0$) this term can still be reintroduced 
in (\ref{metricfinal}). Therefore, the string is always present.

The $h_1$ term corresponds to nothing but a wave along 
the string. There is a solution generating technique developed by 
Garfinkle and Vachaspati \cite{vac, gar} which allows one 
to add such a wave to an already constructed solution
when there is a null Killing vector. This process does not
affect the matter fields. (For a short outline of this method and its 
application
to our case, we refer the reader to the appendix.)

Finally, the $h_2$ part is related to an electromagnetic wave. Note that 
$h_2 \neq 0$ only when the vector field has a radial dependence as can be 
seen from (\ref{vec3}), (\ref{vec4}), (\ref{ozg1}) and (\ref{h0}). 
One may think that 
when $h_0=h_1=0$, a solution with 
$h_2 \neq 0$  exists only for $\e=-1$ since then $H^2$ becomes negative, 
but in fact this is not necessary. In this case 
it can be seen from
(\ref{metric}) and 
(\ref{fh}) that $t$ is no longer the timelike 
coordinate yet the metric is still well-defined.

Let us now analyze the singularity structure of our solution. 
There are three curvature invariants in $D=3$:
\bea
g^{\m\n}R_{\m\n} &=& -8m^2C^2(3C^2-8a^2|S|^2) \, , \nn \\
R^{\m\n}R_{\m\n} &=& 64 m^4 C^4(3C^4 - 16 a^2C^2|S|^2 + 24 a^4|S|^4) \, , 
\la{invariant} \\
\frac{\det (R_{\m\n})}{\sqrt{-g}} &=& 
512m^6C^6(C^2-2a^2|S|^2)^2(C^2 - 4a^2 |S|^2) \, . \nn
\eea

One should also remember that in $D=3$, the Riemann tensor is completely 
determined in terms of the Ricci tensor. Now we proceed with our 
investigation of $\e=-1$ and $\e=1$ cases separately.

\subsubsection{Non-compact Sigma Manifold ($\e=-1$)}

It is easy to see from (\ref{invariant}) that a curvature singularity 
appears as $C^2 \ra \infty$ which implies $|\f| \ra 1$ from (\ref{algrel}). 
To find out whether there is 
any horizon, we define a new radial coordinate
\be
\r=\frac{1}{C^2-1} \, , \hs{5} 0\leq \r < \infty.
\ee
Now the metric (\ref{metricfinal}) becomes
\be
ds^2=-2pf_0 \r^{\frac{1}{2a^2}}dvdt + H^2dv^2 + \frac{d\r^2}{64m^2a^4(\r 
+1)^2} \, ,
\ee
where 
\be
H^2=h_0\r^{\frac{1}{2a^2}} + 
h_1F(\frac{1}{2a^2},1;1+\frac{1}{2a^2};-\frac{1}{\r})
+ h_2 \frac{\r}{1+\r} \, . 
\ee

An inspection of the zeros of $g^{\r \r}=-f_0^2 \r^{1/a^2}$ shows that 
there is no horizon and we have a naked singularity at $\r=0$ 
(or $C^2 \ra \infty$). 

From the curvature scalar (\ref{invariant}) and the Ricci 
tensor (\ref{e1}), it is observed that when $C^2 \ra 1$, i.e. $|\f| \ra 
0$, the solution becomes locally $AdS_3$. In fact when the waves are 
present the metric corresponds 
to a 
generalized Kaigorodov metric \cite{kaig} as was studied in \cite{pope}; 
otherwise the asymptotic geometry is $AdS_3$ \cite{ads2}. From 
(\ref{vec3}) and 
(\ref{vec4}),
one can see that the vector field becomes constant
both at the singularity and at the asymptotic region.

\subsubsection{Compact Sigma Manifold ($\e=1$)}

Let us now define a new radial coordinate
\be
\r=\frac{M}{C^2} \, , \hs{5} M\leq \r < \infty \, ,
\ee
where $M$ is a positive constant. Now the metric (\ref{metricfinal}) 
becomes
\be
ds^2=-2pf_0\left(1-\frac{M}{\r}\right)^{\frac{1}{2a^2}}dvdt
+H^2dv^2 + \frac{d\r^2}{64m^2a^4(\r-M)^2} \,,
\la{metricgeo}
\ee
where
\be
H^2= h_0\left(1-\frac{M}{\r}\right)^{\frac{1}{2a^2}}
+ h_1 F(-\frac{1}{2a^2},1;1-\frac{1}{2a^2};1-\frac{M}{\r})
+ h_2\frac{\r}{M} \, .
\ee

We see that there is a horizon at $\r=M$ (or $C=1$). As $\r \ra M$, 
the curvature scalar (\ref{invariant}) becomes constant and 
the geometry is observed to be locally $AdS$ which is a Kaigorodov 
\cite{kaig}
type of space if the waves are present. When there is only the string, 
the near horizon geometry is $AdS_3$ \cite{ads2}.
In this limit $H^2 \ra h_1+h_2$ and the vector field (\ref{vec3}),
(\ref{vec4}) also becomes constant.

As above there is a curvature singularity as $C^2 \ra \infty$, 
i.e. $\r \ra 0$, but we know from (\ref{algrel}) 
that $C \leq 1$ and therefore the singularity is not accessible. 
Additionally, when $h_2=0$, for $C^2 
\ra 0$ (or $\r \ra \infty$) the solution is asymptotically flat. 
However, even when $h_2 \neq 0$ in the asymptotic regime,
all curvature invariants (\ref{invariant}) still vanish 
and the metric becomes 
\be
ds^2 \ra -2pf_0 dvdt + \frac{h_2}{M}\r dv^2 + \frac{d\r^2}{64m^2a^4\r^2} 
\, ,
\ee
which describes a pp-wave geometry. Therefore the solution with $h_2$ 
term present  
might be interpreted as a string in a space filled with electromagnetic 
radiation in the asymptotic regime. Note that 
in this limit the vector field (\ref{vec3}), (\ref{vec4}) diverges.

Now let us look at the behavior of the geodesics. The geodesic 
equation associated with the metric (\ref{metricgeo}) is:
\be
\frac{1}{64m^2a^4}\left(\frac{\dot{\r}}{\r}\right)^2=
\a \left(1-\frac{M}{\r}\right)^2
-2EP\left(1-\frac{M}{\r}\right)^{2-\frac{1}{2a^2}} + 
E^2H^2\left(1-\frac{M}{\r}\right)^{2-\frac{1}{a^2}}  \, ,
\la{geodesic}
\ee
where the dot denotes derivative with respect to an affine parameter and
$\a=0$ or $\a=-1$ for null or timelike geodesics, respectively. In this
equation $E$ and $P$ are the conserved quantities associated with the flow
of the tangent vector of a geodesic corresponding to $t$ and $v$
variables. 

When $h_2 \neq 0$, the right-hand side of (\ref{geodesic})  
becomes negative as $\r \ra \infty$ since $h_2 < 0$ by (\ref{h0}). This
means that neither timelike nor null geodesics can reach the asymptotic
region. Now lets assume that $h_2=0$. The $h_1$ term is well-defined only
when $1/a^2$ is an odd integer (remember that $1/a^2$ is an integer 
when $\e=1$) since the hypergeometric function diverges otherwise. In 
the asymptotic and the near horizon limits, this function behaves as 
\bea
\lim_{\r \ra \infty} 
F(-{1\over 2a^2}, 1; 1- {1\over 
2a^2}; 1 - \frac{M}{\r}) &\approx&-\frac{1}{2a^2}\log \r \, , \nn \\ 
\lim_{\r \ra M}
F(-{1\over 2a^2}, 1; 1- {1\over
2a^2}; 1 - \frac{M}{\r}) &\approx& 1 \, ,
\eea
and in both limits the last term in the right hand side of (\ref{geodesic}) 
is the dominant term if $1/a^2 >1$. If the geodesics are required to be 
able to reach the $\r \ra \infty$ limit, then $h_1$ should be negative. 
However, since the hypergeometric function changes sign as we 
approach to the horizon ($\r \ra M$), there should be a turning point and 
the geodesics never reach the horizon. If $a^2=1$, the geodesics may 
reach the horizon for large enough $(E^2h_1-2EP)>0$, but even in this case 
they can not go beyond the horizon since 
the $h_1$ term does not change sign whereas others do. 
When $h_1=h_2=0$, the timelike geodesics can not reach the horizon. Moreover, 
the null geodesics do not cross the horizon unless $1/a^2$ is a multiple 
of 4. Since the scalar field can not be extended beyond the horizon, 
there exist physically well-defined strings only for 
$(1/a^2=1,2,3$ mod 4) as was shown in\cite{ads2}. Before we close this 
subsection let us note that when $a^2=1/2$,
the metric of the $h_1=h_2=0$ solution is precisely the string 
solution obtained in \cite{hor} using the 
low energy limit of a three dimensional string theory \cite{ads2}.

\subsection{Flat Sigma Model ($a^2=0$)}

To take the $a^2=0$ limit in our model, 
first one has to rescale
$A_\m \ra a^2 A_\m$ and $\f \ra a\f$. Then we have $C\ra 1$, $S \ra 2a 
\f$, 
and one obtains $N=2$, 
$AdS_3$
supergravity with cosmological constant $-2m^2$ coupled to an $R^{2}$
sigma manifold \cite{ads2}. This coincides with the flat sigma 
model 
limit of the $N=2$ theory discussed in \cite{it} as was shown in 
\cite{mc2}.
The Lagrangian (\ref{ba1}) now becomes
\be
{\cal L} = \sqrt{-g}\left( {1\over 4} R
-{1\over 16m}\, {\e^{\m\n\r} \over \sqrt{-g}} A_\m \del_\n A_\r 
- |\del_\m\f|^2 +2m^2 \right)\, ,
\ee
and its fermionic supersymmetry transformations are
\bea
\d \psi_\m &=& \left(\del_\m + {1\over 4} \o_\m{}^{ab} \c_{ab}
-{i\over 2}\,[ i \f \stackrel{\lra}{\del_\m}\f^* + A_\m] \right) 
\ve +m \c_\m 
\ve\ , \w2
\d\l &=& - ( \c^\m \del_\m \f )  \ve\, .
\eea

To find a 1/2 supersymmetric solution, we again choose the same metric 
ansatz 
(\ref{metric}) and use the same form of scalar and vector fields given in 
(\ref{ans1}) and (\ref{ans2}). Now using the projection condition 
(\ref{proj}) in $\d \l =0$, we find
\be
R'=0 \, , \hs{5} G= -{pF \over H} + {k \over n} \, 
\label{a0}
\ee
from which we write
\be
\f = R_0 e^{in\th} e^{ikt} \, ,
\ee
where $R_0$ is a constant.  Equations (\ref{f1}) and (\ref{f2}) obtained
from $\d \psi =0$ are still valid with $C=1$, which gives
\be
F={f_0 e^{-4pmr} \over H} \, , 
\ee
where $f_0$ is a non-zero integration constant. Now our metric can be 
written as
\be
ds^2= -2pf_0e^{-4pmr}dvdt + H^2dv^2 + dr^2 \, ,
\hs{5} v \equiv \th + \frac{k}{n}t \, . \label{meta=0}
\ee
From $\d \psi =0$, we also 
identify the Killing spinor and vector field components as:
\bea
\ve &=&\sqrt{F} e^{ic_1\th}e^{ic_2t} (p+\c_1)\ve_0 \, , \w2
\chi&=& 2c_1 - 2n R_0^2 \, , \w2
\psi&=& 2c_2 - 2k R_0^2\, ,
\eea
where $c_1$ and $c_2$ are real constants.
The vector (\ref{e2}) and the scalar (\ref{e3}) field 
equations are automatically satisfied, but
Einstein's equations (\ref{e1}) impose:
\be
(H^2)' + 4pmH^2 = -8R_0^2n^2 r + c_0 \, ,
\ee
where $c_0$ is a constant. The most general solution of this equation is
\be
H^2= h_0 e^{-4pmr} + h_1 + h_2 r \, , 
\la{most2}
\ee
where $h_0$ and $h_1=(pmc_0+2R_0^2n^2)/4m^2$ are arbitrary real 
constants and we have 
\be
h_2=-2R_0^2n^2p/m \,. 
\la{matterwave}
\ee

From the curvature invariants (\ref{invariant}), we see that the solution 
has constant negative curvature and it is locally $AdS_3$. Furthermore there 
is no curvature singularity. When $h_1=h_2=0$, the metric is 
the $AdS_3$ metric in Poincar\'e coordinates. For this case even when 
$h_0=0$, one can 
still identify $AdS_3$ by a simple coordinate redefinition. 
The $h_1$ term can be
obtained by using the Garfinkle-Vachaspati method \cite{vac, gar} (see 
the appendix) and it
describes a wave in $AdS_3$. Actually the metric with $h_2=0$ has
already been discussed in \cite{pope} and it corresponds to a generalized
Kaigorodov metric \cite{kaig}. It is obtainable from the $AdS_3$ 
metric by an $SL(2,R)$ transformation \cite{real} and its equivalance to 
the 
extreme BTZ black hole \cite{btz} can be shown \cite{pope, real}.
When $h_2 \neq 0$, the constant $h_1$ can be removed 
by a shift in $r$. This spacetime is another pp-wave in $AdS_3$. 
From (\ref{matterwave}) it is clear that it exists only for a non-zero
scalar field. 
Finally, we would like to point out that
our solutions with waves preserve 1/2 supersymmetry similar to those
found in \cite{klemm, lu} for $D=4,5$ $AdS$ gauged supergravities. 

\section{Conclusions}

In this section we would like to make some further remarks about our
results and suggest some open problems. As we have noted before, the waves
alter the global structure and in several instances the Kaigorodov space
\cite{kaig} or some other pp-wave background emerge. This makes our
solutions suitable for understanding the $AdS/CFT$ duality in the infinite
momentum frame \cite{pope, real} and in the BMN type limit \cite{bmn}. For
example there is still no rigorous notion of ADM mass or energy for 
such spacetimes and our solutions might be useful in this respect. Another 
interesting project is to repeat the RG flow analysis of \cite{rg} for the 
charged strings obtained in this paper.

The presence of a horizon for the $\e=1$ case is very suggestive.  
Although the curvature invariants turn out to be finite at the horizon one
has to be cautious about singularities. As was shown in \cite{myers, yang,
ross, real} there may still be infinities such as the tidal forces felt by 
an
infalling observer. It would be nice to make a more detailed
analysis of this and to see whether the no hair conjecture is supported 
or not. For this purpose it may be necessary  to generalize our 
solutions to have non-constant wave profiles. It is not clear whether 
they would still be supersymmetric.

The waves we found are anchored to the string and when $\e=1$ and $h_2=0$
the solution is asymptotically flat. This makes it useful for trying
to explain the Bekenstein-Hawking entropy by counting the BPS
microstates \cite{stro} as was illustrated in \cite{marolf1} for a 
similar set-up. Of 
course to do this, first higher dimensional origin
of our model has to be identified.

The electromagnetic waves that we obtained do not affect the behaviour of
the scalar field. This perhaps hints to a solution generating mechanism
within our model like the Garfinkle-Vachaspati method. Another example 
of such a technique 
can 
be 
found in \cite{has1} where solutions were
constructed by applying a sequence of manipulations called the null
Melvin twist.

Another attractive avenue is to look for new supersymmetric 
solutions of the model. Especially finding black holes would be very 
appealing. The flat sigma model limit ($a^2 \ra 0$) of our theory
is related to the model discussed in \cite{it} where some black hole 
solutions were found. It is therefore quite likely that such a solution 
exists within our model and it would be very interesting to see the effect 
of non-linear sigma manifolds on the solution presented in \cite{it}. We 
hope to report on this issue soon.

\section*{Acknowledgments}
We would like to thank Gary Gibbons, Rahmi G\"{u}ven, Ali Kaya, Chris
Pope and Ergin Sezgin for useful discussions. This work is partially 
supported by the Scientific and Technical Research Council of Turkey.

\appendix

\section{Garfinkle-Vachaspati Method}

In this appendix we would like to give a brief review of the solution 
generating technique developed by Garfinkle and Vachaspati \cite{vac, 
gar} and apply it to our string solution (\ref{stringmetric}). A detailed 
discussion of this method together 
with its extension to various supergravity theories can be found in 
\cite{myers}. Let $g_{ab}$ be a solution of Einstein's equations with some 
matter source
in a given dimension. 
Let the metric $g_{ab}$ have a null, hypersurface
orthogonal Killing vector $\l^{a}$. Then one can find a scalar $\O$ such
that
\be
\nabla_{a} \l_{b} = \l_{[a} \nabla_{b]} \ln{\O} \, .
\ee
Now a new metric $\h{g}_{ab}$ defined by
\be
\h{g}_{ab} \equiv g_{ab} + \O \, \F \, \l_{a} \, \l_{b} 
\la{newmet}
\ee
yields a new solution to the initial theory with the same matter fields
as in the background solution. The scalar $\F$ here satisfies 
\be 
\l^{a} \, \nabla_{a} \F = 0 \, , \qq \nabla^{a} \nabla_{a} \F = 0 \, . 
\la{depend}
\ee
This implies $\l^{a}$ to be a Killing vector also for the new solution 
$\h{g}_{ab}$. Hence, the new solution describes a traveling wave since
any disturbance must propagate without changing its profile at the speed
of light.

In our case we begin from the string metric (\ref{stringmetric}) 
which can easily be put into the form (this is equivalent to starting from 
(\ref{metricfinal}) with $H=0$)
\be
ds^2=-2pf_0|S|^{\e/a^2} dvdt + dr^2 \, , \hs{5} v\equiv \th + g_0 t \, , 
\ee
where we chose $h_0=0$ without any loss of generality. Now 
using its null Killing vector $k^a=(\del/\del t)^a$,
the scalar $\O$ is calculated easily to be $\O=\O_{0} |S|^{-\e/a^2}$,
where $\O_{0}$ is an arbitrary constant. Because of
(\ref{depend}) $\F$ does not depend on $t$ and is a function of $r$ and 
$v$ only. Using 
(\ref{derivative}), one can explicitly find 
$\F(r,v)$ to be
\be
\F = \frac{p \F_{0}(v)}{4m} \, |S|^{-\e/a^2} \, 
F(-{\e \over 2a^2}, 1; 1- {\e \over 2a^2}; \e |S|^2) + \F_{1}(v) \, ,
\ee
where $\F_{0}$ and $\F_{1}$ are arbitrary functions of $v$ and 
they describe 
the profile of the wave. 
The difference between the new and the old metric (\ref{newmet}) is 
exactly the $h_1$ term in $H^2$ (\ref{most}) where the wave profile 
is fixed to be a constant. This proves our 
identification of the $h_1$ term as a string wave.
Repeating similar 
steps for the $a^2=0$ case one again finds the $h_1$ term in 
(\ref{most2}).

\end{document}